\documentstyle{article}
\newcommand{\rf}[1]{(\ref{#1})}
\renewcommand{\baselinestretch}{1.3}
\newcommand{\be}{\begin{equation}}
\newcommand{\ee}{\end{equation}}

\begin{document}\large
\begin{center}
               THE ENERGY LOSSES OF RELATIVISTIC             \\
         HIGHCHARGED IONS IN THE COLLISIONS WITH ATOMS       \\
                         Matveev V.I.                        \\
         Physics Department, Tashkent State University       \\
                 700095, Tashkent, Uzbekistan                \\
                         Tolmanov S.G.                       \\
    Department of Thermophysics of Uzbek Academy of Science  \\
           Katartal 28, 700135, Tashkent, Uzbekistan         \\
\end{center}
\begin{abstract}
The  energy losses  of heavy  multiplycharged ions
at collisions with light atoms and polarization losses at
moving through matter have been considered  under circumstances
the ion charge $Z\gg 1$ and the relative colliding velocity
$v\gg 1$, so that $Z\sim  v\leq c$ , where c - the light velocity
(atomic units are used). In this region  of  parameters  the
Born approximation is not applicable. The simple formulas
for effective stopping are obtained.The comparison  with
other theoretical results and experiments are given.
\end{abstract}
\section{Introduction}

   Usually an energy loss of a  relativistic charged particle
in collisions with an atom is calculated in the Born approximation
which is applicable if $Z/v\ll1$ (here Z - the ion charge, v - relative
velocity of collision, atomic units are used) (Berestetskii et al 1989).
However, there  were  some
resent experiments which use ions of such a  large  charge  so
the above mentioned condition was  violated  even  at
$v\leq c\approx137$ (see for instance Kelbch et al 1986, Berg et al 1988,
Berg et al 1992, Scheidenberger et al 1994).
The approximations applicable at $Z/v\sim1$ - Eikonal
approximation and its modifications (McGuire 1982, Crothers and McCann 1983),
the
Method of Sudden Perturbations (Eichler 1977, Salop and Eichler 1979,
Yudin 1981, Yudin 1991),
as  well  as the Method of the Classical Trajectories
(Olson 1988) demand
considerable numerical calculations  even  in  nonrelativistic
range of collision velocities. Relativistic collision velocities make the
the calculations even more complicated (Yudin 1991).

In the present paper we consider the energy loss  of  heavy
relativistic highcharged ions in collisions with  light
(nonrelativistic) atoms and polarizations losses at moving through
the matter under conditions $Z\sim v\leq c$, $Z\gg1$,
$v\gg1$  by following a simple approach
(Matveev 1991) and its  relativistic modification
(Matveev and Musakhanov 1994). We  are  going
to obtain simple formulas describing an effective stopping and
compare it's results with the results of other theoretical methods and
experiments.

\section{Collisions with separated atom}

Accordingly to (Landau and Lifshitz 1974) average losses of
energy in collisions are characterized by the effective
stopping
\be
\kappa=\sum_{n}(\epsilon_{n}-\epsilon_{0})\sigma_{n}\;.
\label{equ1}\ee
Here $\epsilon_{n}$  and $\epsilon_{0}$ are the energies of
excited  $|n>$  and  bound  $|0>$
atom's states, $\sigma_{n}$ is the cross section for excitation  of  state
$|n>$. First, we consider, for simplicity, the collision of a
relativistic highcharged ion with a hydrogen  atom.  Accordingly
to (Matveev 1991, Matveev and Musakhanov 1994) the whole
interval $0<b<\infty$ of all possible values of impact parameter  b
may be divided into three ranges:
\be
 A)\,\,0<b<b_{1}\,;\;\;B)\,\,b_{1}<b<b_{0}\,;\;\;C)\,\,b_{0}<b<\infty\, ,
\label{equ2}\ee
which correspondent to small, middle and large values of an
impact parameter. The border amounts $b_{1},\;b_{0}$  are
(Matveev and Musakhanov 1994)
$$ b_{1}\sim1\,;\;\;b_{0}\sim\ v/\sqrt{1-\beta^{2}}\,;\;\;\beta=v/c\,.$$

   We are going to calculate $\kappa$ in each  range  from  \rf{equ2}  and
obtain total effective stopping by summing those values .  The
exact meaning of $b_{0}$ and $b_{1}$ are  unimportant  for  us,  since
dependence $\kappa$  upon  sewing  parameters  $b_{1},\;b_{0}$ appears to  be
logarithmic, which lead to the correct sewing of contributions
from each region \rf{equ2} and vanishing any dependence $\kappa$ on $b_{1}$ and
$b_{0}$  in the final result.

   The range of small impact  parameters  is  the  range  of
large momentum transfers, thus we needn't  take  into  account
the interaction of an electron with an atom nuclei, and
therefore we can consider the scattering of an ion on  an  electron
having rested before collision (Berestetskii et al 1989 \S82).  In
this case an effective stopping may be expressed through using
$\sigma(\epsilon)$ which is a cross section  for  collision
with energy transfer $\epsilon$:
\be
\kappa=\int^{\epsilon_{max}}_{\epsilon_{min}}\!\!\epsilon
\sigma(\epsilon)d\epsilon\;\;.
\label{equ3}\ee
Here we can't, unlike (Berestetskii et al 1989),  use $\sigma(\epsilon)$  in
the Born approximation, because a condition $Z/v\ll1$ is  invalid.
To go on, we assume colliding ion to be  an  infinitely  heavy
particle, and consequently it doesn't change its moving.  This
case the cross section for scattering of an ion moving at fixed
velocity on the free and motionless electron may  be  obtained
by transformation to the system where ion  is  motionless.  We
mark $\theta$ a respective angle of scattering. Following to
(Berestetskii et al 1989) we can maintain an energy transfer to be a
function $\epsilon=\epsilon(\theta)$, except the case of a superhigh
energy. That function has a form
\be
\epsilon(\theta)=\frac{2v^{2}}{(1-\beta^{2})}Sin^{2}\frac{\theta}{2}\;\;.
\ee
The values $\epsilon_{max}$ and $\epsilon_{min}$ from \rf{equ3} are reached
at $\theta=\pi$ and $\theta=\theta_{min}$ respectively.
As a result we can rewrite \rf{equ3} in a form
\be
\kappa=2\pi\frac{Z^{2}}{v^{2}}\int^{\pi}_{\theta_{min}}\!\frac{\sigma(\theta)}{\sigma_{R}(\theta)}\,ctg{\frac{\theta}{2}}d\theta\;\;,
\label{equ5}\ee
Here $\sigma(\theta)$ is the cross section for the scattering of electron
on motionless ion with a charge Z (at electron velocity v),
which is obtained (Ahieser and Berestetskii 1969, Mott and Massey 1965)
from exact solution of a scattering  problem  for  the  Dirac equation;
$\sigma_{R}$ is the Rutherford cross section:
\be
\sigma_{R}=\frac{Z^{2}(1-\beta^{2})}{c^{4}\beta^{4}(1-cos{\theta})^{2}}\;\:.
\ee
The ratio $\sigma(\theta)/\sigma_{R}(\theta)\!\rightarrow\!1$ at
$\theta\!\rightarrow\!0$
(Ahieser and Berestetskii 1969, Doggett and Spenser 1956), and hence in order
to define $\theta_{min}$
we can use the quassiclassical connection (Landau and Lifshitz 1973)
between a scattering angle and an impact parameter
(see also the obvious qualitative picture of collision suggested in
(Matveev 1991))
\be
\theta_{min}=\frac{2Z}{v^{2}b_{1}}\sqrt{1-\beta^{2}}\;.
\ee
One can easy see that the integral \rf{equ5} depends on angle
$\theta_{min}$ in logarithmic way (at small $\theta_{min}$), thus the formula
\rf{equ5} may be
written in a form:
\be
\kappa=4\pi\frac{Z^{2}}{v^{2}}\ln{\frac{v^{2}b_{1}}{Z\sqrt{1-\beta^{2}}\,a(Z,v)}}\:.
\label{equ8}\ee
The function a(Z,v) are defined by us from comparing \rf{equ8}
with the numerical calculation results by formula \rf{equ5}
with the ratio $\sigma/\sigma_{R}$ from (Doggett and Spenser 1956).
As a result we can approximate function a(Z,v) with a following formula:
\be
  a(Z,v)=(-0.23016\cdot\alpha-(1.00832\cdot\alpha-0.32388)\beta^{2}+1)^{2}\;,
\label{equ9}\ee
here $\alpha=Z/c$.
In the tables number 1 and number  2  the  effective  stopping
calculated (at $b_{1}=1$) by formula \rf{equ5} (table 1) and  by  formula
\rf{equ8} with substituting the function a(Z,v) from \rf{equ9} (table 2) are
brought. In these tables the data  of  1st  column  are  ion's
energies (in Mev/nucleon) correspondent to the relative velocities from
(Doggett and Spenser 1956); the data  of  next  columns  are
$\kappa$ values (in atomic units) for the ions with a  charge:  6,  13,
29, 50, 82, 92 respectively. One can see from the tables  that
the supposed approximation is enough good, at any rate in  the
limits of v and Z which the data of (Doggett and Spenser 1956)
are brought for. Besides we have to  emphasize  that  function
$a(Z,v)\rightarrow1$ at $\beta\rightarrow0$ (nonrelativistic limit) and
$\alpha\rightarrow0$,
which  conforms to the fact that $\sigma/\sigma_{R}\rightarrow1$ at
$\beta\rightarrow0$,
$\alpha\rightarrow0$.
\begin{table}
\begin{tabular}{|c|c|c|c|c|c|c|}                                    \hline
Ion's energy & \multicolumn{6}{c|}{Ion's charges}                \\ \cline{2-7}
 Mev/nucleon &  6    &  13    & 29    & 50    &  82    & 92      \\ \hline
  91.8       & 0.894 & 3.714  &16.247 &44.368 &110.067 &135.497  \\ \hline
  183.6      & 0.554 & 2.333  &10.411 &29.185 &75.125  &93.469   \\ \hline
  367.2      & 0.372 & 1.580  & 7.160 &20.452 &53.402  &68.540   \\ \hline
  734.4      & 0.280 & 1.200  & 5.495 &15.893 &43.418  &55.109   \\ \hline
  1285.2     & 0.245 & 1.056  & 4.872 &14.181 &39.220  &49.993   \\ \hline
  1836       & 0.235 & 1.015  & 4.697 &13.705 &38.063  &48.618   \\ \hline
  3672       & 0.231 & 1.003  & 4.670 &13.671 &38.079  &48.701   \\ \hline
  7344       & 0.239 & 1.043  & 4.879 &14.310 &40.246  &50.941   \\ \hline
  18360      & 0.257 & 1.129  & 5.311 &15.595 &43.308  &55.302   \\ \hline
\end{tabular}
\caption{Effective stoping (at.units) - result of numerical integration
by formula (5) with using  $\sigma/\sigma_{R}$  from (Doggett and Spenser
1956)}
\vskip 1cm
\begin{tabular}{|c|c|c|c|c|c|c|} \hline
Ion's energy & \multicolumn{6}{c|}{Ion's charges}                  \\
\cline{2-7}
Mev/nucleon  &  6    &  13    & 29    & 50    &  82    & 92       \\ \hline
  91.8       & 0.891 & 3.697  &16.056 &43.666 &110.134 & 137.246   \\ \hline
  183.6      & 0.552 & 2.319  &10.268 &28.454 &73.567  & 92.369    \\ \hline
  367.2      & 0.370 & 1.570  & 7.066 &19.922 &52.838  & 66.916    \\ \hline
  734.4      & 0.278 & 1.192  & 5.439 &15.559 &42.255  & 53.985    \\ \hline
  1285.2     & 0.244 & 1.051  & 4.834 &13.951 &38.430  & 49.379    \\ \hline
  1836       & 0.234 & 1.010  & 4.666 &13.520 &37.462  & 48.252    \\ \hline
  3672       & 0.230 & 0.999  & 4.647 &13.529 &37.685  & 48.636    \\ \hline
  7344       & 0.238 & 1.040  & 4.860 &14.185 &39.523  & 50.991    \\ \hline
  18360      & 0.256 & 1.125  & 5.291 &15.475 &43.018  & 55.405    \\ \hline
\end{tabular}
\caption{Effective stoping (at.units), by formulas  (8) with using
a funstion a(Z,v) defined from (9)}
\end{table}

   The collisions with the middle impact parameter $b_{1}<b<b_{0}$ make
the main contribution to the cross sections for the inelastic
process (Matveev 1991, Matveev and Musakhanov 1994). The
energy transfer in that collision is $\epsilon\leq1\sim$ ionization potential
of an atom, and therefore the contribution to the effective
stopping  from that collision can't be accounted  by using a
perturbation theory (Eichler 1977, Yudin 1981, Matveev 1991,
Matveev and Musakhanov 1994).
We must note else, that an atom
electron  is nonrelativistic as before so after collision with
an impact parameter from range $b_{1}<b<b_{0}$
(Matveev and Musakhanov 1994). The contribution of a collision with such impact
parameter to an effective stopping may be easy obtained from
formula \rf{equ1} by substituting to it the cross section for the
inelastic process (Matveev and Musakhanov 1994)
\be
\sigma_{n}=\int^{b_{0}}_{b_{1}}\!\!2\pi b\mid<n\mid
exp(-i\vec{q}\vec{r})\mid0>\mid^{2}db\:,
\label{equ10}\ee
Here $\vec{q}=2Z\vec{b}/(vb^{2})$ and $\vec{b}$ is
impact parameter vector, $\vec{r}$ is a coordinate of atomic electron.
On repeating the calculations of (Landau and Lifshitz 1974)
(but our case is simpler, because the upper limit in
integration \rf{equ10} doesn't depend on the final state of
an atom) we have for an effective stopping:
\be
\kappa=\sum_{n}(\epsilon_{n}-\epsilon_{0})\sigma_{n}=4\pi\frac{Z^{2}}{v^{2}}\ln{\frac{q_{1}}{q_{0}}}\:,
\label{equ11}\ee
here $q_{0}=2Z/(vb_{0}\:);\;\;\;q_{1}=2Z/(vb_{1})$ .

   We need to account the contribution from collisions with an
impact parameter belonging to the range $b_{0}<b<\infty$. In  this
range the interaction between an  ion  and  an  atom  may  be
accounted  by  using  a  perturbation  theory  (Eichler 1977,
Matveev 1991, Matveev and Musakhanov 1994).
Amplitude for  the
transition of an atom from a state $|0>$ to a state $|n>$  may  be
obtained following to (Moiseiwitsch 1985)
\be
A_{n}=\frac{2iZ}{v^{2}}\Omega_{n}\vec{r}_{0n} \left[
i\frac{\vec{v}}{v}(1-\beta^{2})K_{0}(\xi)+\frac{\vec{b}}{b}\sqrt{1-\beta^{2}}\,K_{1}(\xi)\right]\:,
\label{equ12}\ee
here
$\;\Omega_{n}=\epsilon_{n}-\epsilon_{0}\:;\;\:\xi=\Omega_{n}b\sqrt{1-\beta^{2}}/v\:$; $K_{0}(\xi)$,
$K_{1}(\xi)$ - the McDonald functions,
$\vec{r}_{0n}=<0|\vec{r}|n>$.

 The cross section correspondent to \rf{equ12} is
$$\sigma_{n}(b_{0}<b<\infty)=\int \!d^{2}b\mid A_{0n}\mid^{2}\;.$$
The integrating in this formula is made in limits:
the angle of vector $\vec{b}$ changes from 0 to $2\pi$ and $b_{0}<b<\infty$.
In a result the cross section has a form
\footnote{To say rigorously, formula \rf{equ13} was obtained from supposition
that
 $\xi=\Omega_{n}b\sqrt{1-\beta^{2}}/v\ll1$ ($\Omega_{n}\sim 1$), then
$b\ll v/\sqrt{1-\beta^{2}}\sim b_{0}$, so following sewing may be performed
exactly at such b.}:
\be
\sigma_{n}=4\pi\frac{Z^{2}}{v^{2}} \mid x_{0n}\mid^{2} \left(
\ln{\frac{4v^{2}}{\eta^{2}b_{0}^{2}\Omega_{n}^{2}(1-\beta^{2})}} -
\beta^{2}\right)\:,
\label{equ13}\ee
Here $\eta=e^{B}=1.781$ (B=0.5772 - the Euler constant), $x_{0n}=<n|x|0>$.
The contribution to a effective stopping from collisions with
large impact parameters may be calculated through substituting
expression \rf{equ13} to formula \rf{equ1}
\be
\kappa=4\pi\frac{Z^{2}}{v^{2}}\left\{ \ln{\frac{2v}{\eta I
b_{0}\sqrt{1-\beta^{2}}}} - \beta^{2}/2 \right\}\:,
\label{equ14}\ee
Here we have introduced "average atom energy"- I
(Berestetskii et al 1989 \S 82), which is
\be
\ln{I}=\frac{\sum_{n}(\epsilon_{n}-\epsilon_{0})\mid x_{0n}\mid^{2}
\ln{(\epsilon_{n}-\epsilon_{0})}}
{\sum_{n}(\epsilon_{n}-\epsilon_{0})\mid x_{0n}\mid^{2}}
\label{equ15}\ee

   The total effective stopping of a relativistic highcharged
ion on a hydrogen atom is obtained on summing the results of
formulae \rf{equ14},\rf{equ11} and \rf{equ8}:
\be
\kappa=4\pi\frac{Z^{2}}{v^{2}}\left\{ \ln{\frac{2v^{3}}{\eta I
Z(1-\beta^{2})a(Z,v)}} - \beta^{2}/2 \right\}\:.
\label{equ16}\ee
We would like to bring $\kappa$ value calculated in the Born
approximation from (Berestetskii et al 1989)
\be
\kappa=4\pi\frac{Z^{2}}{v^{2}}\left\{ \ln{\frac{2v^{2}}{I(1-\beta^{2})}} -
\beta^{2} \right\}.
\label{equ17}\ee

   It  should  be  noticed  that  nonrelativistic  limit   for
ionization losses \rf{equ16} (in accounting of
$a(Z,v)\rightarrow1$ at $\beta\rightarrow0$ and $\alpha\rightarrow0$)
\be
\kappa=4\pi\frac{Z^{2}}{v^{2}}\ln{\frac{2v^{3}}{I\eta Z}}
\label{equ18}\ee
has a form which is similar to  wellknown  Bohr formula
(Bohr 1913) obtained from classical representations. Besides
we have to notice that our result \rf{equ16} obtained on the  basis
of approach (Matveev 1991, Matveev and Musakhanov 1994),
which is valid at $Z\sim v\gg1$ only, doesn't have
a way to pass (so does the Bohr formula) to the Born
approximation which is valid at $v/Z\ll1$.

As followed from our way of obtaining formula \rf{equ16}, in order
to generalize it as well as the Born losses \rf{equ17} to  the  case
of collisions with multielectron atoms (whose  electrons  have
velocities $v_{a}\ll v$, $v$ is ion's velocity) we  need  to  multiply
the right part of formula \rf{equ16} on $Z_{a}$ ($Z_{a}$ is the number of atom
electrons) and change value I for average  atom  potential I$_{a}$
defined, as before, from  \rf{equ15},  but  with  values  $\epsilon_{n}$,
$\epsilon_{0}$ and $\mid x_{0n}\mid^{2}$ calculated for complex atom.

\section{Energy losses in matter}

Let's consider an energy loss of  relativistic  highcharged
ion  moving  through  matter.  This  losses  are  the  sum  of
macroscopic (polarization) losses and losses at collision with
separated atoms. The ion velocity hold to be much bigger  than
typical velocities of atomic electrons or at least the majority of
them. Accordingly to
(Fermi 1940) polarization losses of a charged particle moving
in  the  matter are
defined by the flow of the energy of electromagnetic field of a particle
through the cylinder with radius  $b'_{0}$ built  around a particle
trajectory.  The effective stopping is obtained by
dividing of the flow on a particle velocity:
\be
\kappa=\frac{Z^{2}b'_{0}}{\pi v^{2}}\int^{+\infty}_{-\infty}\!\!d\omega
K_{0}(b'_{0}\xi)K_{1}(b'_{0}\xi^{*})\left(\frac{1}{\varepsilon(\omega)}-\beta^{2}\right) i\omega\xi^{*}\:,
\ee
here $\xi^{2}=\omega^{2}(v^{-2}-c^{-2}\varepsilon(\omega))$,
$\varepsilon(\omega)$ -
dielectrical penetration coefficient.
At enough small cylinder radius $b'_{0}$, i.e.
\be
\mid b'_{0}\xi\mid\ll1
\label{equ20}\ee
we have
\be
\kappa=\frac{i Z^{2}}{\pi v^{2}}\int^{+\infty}_{-\infty}\!\!\omega d\omega
\left\{ \frac{1}{\varepsilon(\omega)} -
\beta^{2}\right\}\ln{\frac{2}{\eta b'_{0}\xi}}\:,
\label{equ21}\ee
where $\eta$=1.781 (like \rf{equ13}).
Other side, accordingly to (Landau and Lifshitz 1982)
energy  loss  may  be counted as a work of electromagnetic field
per unit of way
\be
\kappa=\frac{i Z^{2}}{\pi}\int^{q_{0}}_{0}\!\int^{+\infty}_{-\infty} \! \omega
d\omega qdq
\frac{(v^{-2}-c^{-2}\varepsilon(\omega))}{\varepsilon(\omega)(q^{2}+\xi^{2})}\:.
\label{equ22}\ee
Formula \rf{equ21} will coincide with \rf{equ22} if  condition  \rf{equ20}  is
valid and $q_{0}=2/(\eta b'_{0})$: surely,
\be
\ln{\frac{2}{\eta\xi
b'_{0}}}\approx\frac{1}{2}\ln\left(\frac{2^{2}}{\eta^{2}\xi^{2}
b'_{0}{}^{2}}+1\right)=\int^{q_{0}}_{0}\!\frac{qdq}{q^{2}+\xi^{2}}\:.
\label{equ23}\ee

There are two distinguished cases to be considered separately:
a) $v^{2}<c^{2}/\varepsilon_{0}$
($\varepsilon_{0}=\varepsilon(0)$) - dielectrical penetration coefficient
in static field) and b) $v^{2}>c^{2}/\varepsilon_{0}$.
In first case (Landau and Lifshitz 1982):
\be
\kappa=4\pi\frac{N Z^{2}}{v^{2}}\left[
\ln{\frac{q_{0}v}{\bar{\omega}\sqrt{1-\beta^{2}}}} - \frac{\beta^{2}}{2}
\right]\:\:,q_{0}=2/(\eta b'_{0})\:.
\ee
Here $\bar{\omega}$ is a average amount of frequency of atom's electrons
motion:
$$
\ln{\bar{\omega}}=\frac{\int^{\infty}_{0}\!\omega\eta''(\omega)\ln{\omega}\,d\omega}{\int^{\infty}_{0}\!\omega\eta''(\omega)d\omega}\:,
$$
here $\eta''=Im(\varepsilon^{-1}(\omega))$.
In second case ($(v^{2}>c^{2}/\varepsilon_{0})$): 1) if
the particle  energy  is  not
rather large (motion energy is either less than or order of a rest energy of
ion)
we can use the formula  \rf{equ23};
2) In  ultrarelativistic  case
(Landau and Lifshitz 1982)
\be
\kappa=2\pi\frac{N Z^{2}}{c^{2}}\ln{\frac{q_{0}^{2}c^{2}}{4\pi N}}\:.
\ee

Now we must sum the macroscopic losses and the energy losses on
separated atoms. In order to do it rewrite the condition  \rf{equ20}
in a form:
$$
b'_{0}\ll
v/\mid\omega\sqrt{1-\beta^{2}\varepsilon}\mid<v/\sqrt{1-\beta^{2}}\sim b_{0}\:,
$$
where, for estimations, was assumed that $\omega\sim \omega_{a}\sim1$ -
typical atomic frequency.
The condition of macroscopic method's being applicable is $b'_{0}\gg
b_{1}\sim1$ -
typical atomic size. Thus the lowest border $b'_{0}$ is between $b_{1}$ and
$b_{0}$:
\be
b_{1}\ll b'_{0}\ll b_{0}
\label{equ26}\ee
On comparing this condition with \rf{equ2} we conclude that in order
to  obtain  the  total   energy   losses   of   relativistic
highcharged ion moving through the matter we have to  sum  the
contribution from ranges A and B from  \rf{equ2}  (in  range  B  the
upper limit is equal to $b'_{0}$) and polarization loss.
The sum of contribution from ranges A and B according  to  \rf{equ8}
and \rf{equ11} is equal
\be
\kappa=4\pi\frac{Z^{2}N}{v^{2}}\ln{\frac{v^{2}b'_{0}}{Z\,a(Z,v)\sqrt{1-\beta^{2}}}}\:.
\ee
We  changed  the  number  of  atom  electrons  -  Z$_{a}$ for  the
number of electron in the unit of volume - N (as was hold in
Landau and Lifsitz 1982).
By summing \rf{equ23} and \rf{equ26} we obtain the total energy loss
of relativistic highcharged ion moving through  matter  for
the case
$v^{2}<c^{2}/\varepsilon_{0}$
\be
\kappa=4\pi\frac{Z^{2}N}{v^{2}}\left[\ln{\frac{2v^{3}}{Z\eta(1-\beta^{2})a(Z,v)\bar{\omega}}}- \beta^{2}/2 \right]\:.
\ee
As was said this formula is used often for the  case
$v^{2}>c^{2}/\varepsilon_{0}$
and for the particle whose velocity is not too high.  It
should be noticed that \rf{equ22} is different from \rf{equ18}  describing
the energy losses on separated  atoms  with  changing  average
potential I for $\bar{\omega}$ only (there is the same situation in
Landau and Lifshitz 1982).
In ultrarelativistic case on doing the same way as  before  we
obtain total effective stopping in a form:
\be
\kappa=2\pi\frac{Z^{2}N}{c^{2}}\left[\ln{\frac{4c^{6}}{Z^{2}\eta^{2}(1-\beta^{2})^{2}a^{2}(Z,v)4\pi N}}-1/2\right]\:.
\label{equ29}\ee
One can see that in this case accounting the  polarization  loss
leads to slower growing an effective stopping power
with increasing a collision velocity, in comparing with case of losses on
separated atoms \rf{equ18}.
One may rewrite the formula \rf{equ29} introducing the Fermi corrections
on density effect:
\be
\kappa=4\pi\frac{Z^{2}N}{v^{2}}\left[\ln{\frac{2v^{3}}{Z\eta(1-\beta^{2})a(Z,v)I_{a}}}- \beta^{2}/2-\delta/2 \right]\:,
\ee
here the meaning of average ionization potential $I_{a}$ and the Fermi
corrections $\delta/2$ are brought from (Sternheimer et al 1984,
Inokuti and Smith 1982).
\begin{table}
\begin{tabular}{|lccccc|} \hline
projectile              &target&\multicolumn{2}{c}{calculation (*)}&Our results
& experiment (*) \\
\hline
$^{86}_{36}$Kr          &\,    &   \,   &  \,     &        &   \,            \\
900 MeV/u               &Be    & 2.346  & 2.438   &2.55200 & 2.432$\pm$0.037 \\
$(\beta=0.861)$         &\,    &   \,   &  \,     &        &   \,            \\
\hline
$^{136}_{\enspace54}$Xe &Be    & 5.488  & 5.812   &5.86187 & 5.861$\pm$0.076 \\
780 MeV/u               &C     & 6.014  & 6.378   &6.43478 & 6.524$\pm$0.084 \\
$(\beta=0.839)$         &Al    & 5.404  & 5.755   &5.80985 & 5.806$\pm$0.121 \\
       \,               &Cu    & 4.703  & 5.036   &5.08741 & 5.077$\pm$0.066 \\
       \,               &Pb    & 3.654  & 3.942   &3.98736 & 3.959$\pm$0.063 \\
\hline
\end{tabular}
\caption{Experimental and theoretical meanings of energy losses (in
$Mev/(mg/cm^{2})$)
for various combinations target-ion.\newline (*)-Scheidenberger et al 1994}
\end{table}

In table 3 the theoretical and experimental meaning of
effective stopping \linebreak (in  $Mev/(mg/cm^{2})$) for various ions
(we limited ourselves with Kr and Xe which have enough large charge)
and targets from (Scheidenberger et al 1994), as well as our results are
brought:
the column 1 - ion, it's energy  (in   Mev/nucleon)  and
amount $\beta=v/c$; the column 2 - the target's nature;
the column 3 - the results of calculations (Scheidenberger et al 1994) by the
Bete formula
with the Fermi corrections; the column 4 -
the results of calculating (Scheidenberger et al 1994)
by the Bete formula with accounting Fermi corrections, Mott corrections
(Mott 1929) and Bloch corrections (Bloch 1933);
column 5 - are our results; column 6 - the experimental meaning for
effective stopping (Scheidenberger et al 1994).
One can see that our results
are in rather good accordance with experiment.

\section{Conclusion}

The simple approach suggested in the papers (Matveev 1991, Matveev and
Musakhanov 1994)
enable us  to  estimate  an  effective   stopping   of   relativistic
highcharged ion in collisions  with  separated  atoms  and  at
motion in matter for many practically important  cases, since the
formulas obtained in this paper enable us to use the wellknown
method of phenomenological corrections usually used in applied calculation.
The region of applicability of our formulas $Z\sim v\leq c$
doesn't permit to make direct transition to the Born approximation
($Z/v\ll 1$). But that fact is not too significant limitation,
because the region of applicability of the Born approximation for the ion
with enough large charge (for example $Z=92$) is unreachable even at
$v\approx c$.

The authors acknowledge Dr. Cristoph Scheidenberger, GSI, Darmstadt, Germany
for attention to our work and information about his results.
\newpage
\centerline{References}
\begin{itemize}
\item[{}]{\hskip -1cm Ahieser A.I. and Berestetskii V.B. 1969, Quantum
Electrodynamics. (Moscow: Nauka).}
\item[{}]{\hskip -1cm Berg H.,Dorner R., Kelbch C.,Kelbch S., Ullrich J.,
Hagmann S.,
 Richard  P.,  Schmidt-Borking H., Schlachter A.S., Prior M.,
 Crawford H.J., Engelage J.M., Flores I., Loyd D.H., Pedersen J.
 and Olson R.E.  1988, J.Phys.B: At.Mol.Opt.Phys.\underline{21}, 3929-39.}
\item[{}]{\hskip -1cm Berg H., Ullrich J., Bernstein E., Unverzagt M.,
Spielberger L.,
 Euler J., Schardt D., Jagutzki O., Schmidt-Borking H., Mann R.,
 Mokler P.H., Hagmann S. and Fainstein P.D.  1992, J.Phys.B: At.
 Mol.Opt.Phys. \underline{25}, 3655-70.}
\item[{}]{\hskip -1cm Berestetskii V.B., Lifshitz  E.M.  and  Pitaevskii  L.P.
 1989,
 Quantum Electrodynamics. (Moscow: Nauka).}
\item[{}]{\hskip -1cm Bohr N. 1913 Phil. Mag. \underline{25}, 10-31.}
\item[{}]{\hskip -1cm Bloch F., 1933 Ann.der Phys., \underline{16}, p.285.}
\item[{}]{\hskip -1cm Crothers D.S.F. and McCann J.U.  1983, J.Phys.B:
At.Mol.Opt.Phys.
\underline{16}, 3229-42.}
\item[{}]{\hskip -1cm Doggett J.A. and Spencer L.V.  1956, Phys.Rev.
\underline{103}, 1597-601.}
\item[{}]{\hskip -1cm Eichler J.H.  1977, Phys.Rev.A \underline{15}, 1856-62.}
\item[{}]{\hskip -1cm Fermi~E., 1940, Phys.Rev., \underline{57}, p.485}
\item[{}]{\hskip -1cm  Inokuti~M., Smith~D.Y., 1982, Phys.Rev.B,
\underline{25}, p.61.}
\item[{}]{\hskip -1cm Kelbch S., Ullrich J.,Rauch W., Schmidt-Borking H.,
Horbatsch M.,
 Dreizler R.M., Hagmann S., Anholt R., Schlachter A.S., Mtller A.,
 Richard P., Stoller Ch., Cocke C.L., Mann R., Meyerhof W.E. and
 Rasmussen J.D.,  1986, J.Phys.B: At.Mol.Opt.Phys. \underline{19}, L47-L52.}
\item[{}]{\hskip -1cm Landau L.D. and Lifshitz E.M.  1982, Electrodynamics of
Continuous Media, (Moscow: Nauka)}
\item[{}]{\hskip -1cm Landau L.D. and Lifshitz E.M.  1974, Quantum Mechanics,
 (Moscow: Nauka)}
\item[{}]{\hskip -1cm Landau L.D. and Lifshitz E.M.  1973, Clasical Theory of
Fields,
 (Moscow: Nauka)}
\item[{}]{\hskip -1cm Matveev V.I.  1991, J.Phys.B: At.Mol.Opt.Phys.
\underline{24}, 3589-97.}
\item[{}]{\hskip -1cm Matveev V.I.  and  Musakhanov  M.M.   1994,
Zh.Eksp.Teor.Fiz.
\underline{105},  280-87.}
\item[{}]{\hskip -1cm McGuire J.H.  1982, Phys.Rev.A \underline{26}, 143-7.}
\item[{}]{\hskip -1cm Moiseiwitsch B.L.  1985, Phys. Reports, \underline{118},
133-77.}
\item[{}]{\hskip -1cm Mott~N.F. and Massey H.S.W.  1965, The Theory Of Atomic
Collisions. Thrird Edition, Oxford at the Clarendon Press.}
\item[{}]{\hskip -1cm Mott~N.F., 1929, Proc.Roy.Soc.  \underline{A124}, p.425.}
\item[{}]{\hskip -1cm Olson R.E. 1988, In "Electronic and Atomic Collisions"
\linebreak (H.B.Gilbody, W.R.Newell, F.H.Read and A.C.Smith, eds) p.271,
 Elsevire Science, London.}
\item[{}]{\hskip -1cm Salop A. and Eihler  J.H.   1979,  J.Phys.B:
At.Mol.Opt.Phys.\underline{12},
 257-64.}
\item[{}]{\hskip -1cm Scheidenberger~C., Geissel~H., Mikelsen~H.H.,Nickel~F.,
Brohm~T., Folger~H.,
Irnoch~H., Magel~A., Mohar~M.F., Munzenberg~G., Pfutzner~M., Roeckl~E.,
Schall~I., Schardt~D., Schmidt~K.-H., Schwald~W., Steiner~M., Stohler~Th.,
Summerer~K.,Vieera~D.J., Voss~B., Weber~M.,
Phys.Rev.Lett. 1994, \underline{73}, p.50-58.}
\item[{}]{\hskip -1cm Sternheimer~R.M., Berger~M.J. and Seltzer~S.M., 1984,
Atomic Data
    and Nuclear Data Tables, \underline{30}, p.261.}
\item[{}]{\hskip -1cm Yudin G.L.  1981, Zh.Eksp.Teor.Fiz. \underline{80},
1026-37.}
\item[{}]{\hskip -1cm Yudin G.L.  1991, Phys.Rev. \underline{A44}, 7355-60.}
\end{itemize}
\end{document}